\documentclass[amsmath, amssymb,aps, pra,floatfix, twocolumn,mathbbm,superscriptaddress,longbibliography]{revtex4-2}
\usepackage{amsmath}
\usepackage{amssymb}
\usepackage{graphicx,dsfont}
\usepackage[dvipsnames]{xcolor}
\usepackage[colorlinks=true,citecolor=Orange,linkcolor=Orange,urlcolor=Purple]{hyperref}
\usepackage{physics}
\usepackage{braket}

\usepackage{xcolor}

\newcommand\idop{\mathds{1}}

\begin{document}
	\title{On the Significance of Interferometric Revivals for the\\ Fundamental Description of Gravity}
	\author{Kirill Streltsov}
	\author{Julen S. Pedernales}
	\author{Martin B. Plenio}
	\affiliation{Institut f\"ur Theoretische Physik und IQST, Albert-Einstein-Allee 11, Universit\"at Ulm, D-89081 Ulm, Germany}
	\date{\today}
	
	\begin{abstract}
		We show that an interaction between a harmonic oscillator and a two-level test mass (TLTM) mediated by a local operations and classical communication (LOCC) channel produces a signature that in {[D. Carney et al., PRX Quantum 2, 030330 (2021)]} is claimed to be exclusively reserved for channels that can transmit quantum information. We provide an explicit example based on a measurement-and-feedback channel, explain where the proof of Carney et al. fails, discuss to what degree setups of this type can test the nature of the gravitational interaction and remark on some fundamental implications that an LOCC model of gravity may have in black hole physics.
	\end{abstract}

	\maketitle
	
	\section{Introduction}
	The reconciliation of quantum mechanics and gravity is a long-standing open problem in physics, but progress towards a satisfying solution has long been hindered by the inaccessibility of the necessary experimental conditions. The ambition to perform tests that target the question of whether gravity in fact needs to be quantized  \cite{Feynman57, karolyhazy1966, penrose1996, kafri2013} goes back at least as far as $1957$ but has seen a considerable gain in momentum six decades later with the recent remarkable progress in the control of the quantum degrees of freedom of massive objects \cite{gonzalez-ballestero2021,Aspelmeyer2014,delic2020,streltsov2021}. 
	
	Proposed tests aim at detecting modifications to the unitary evolution predicted by quantum mechanics \cite{pikovski2012, kafri2014, albrecht2014, kumar2020} or ask whether gravity can entangle two parties as this would falsify the assumption of a classical force carrier and thereby conclude the non-classical nature of the gravitational interaction \cite{kafri2013,krisnanda2017}.
	Proposals that aim to realize such tests include  \cite{kafri2013,bose2017,krisnanda2020,pedernales2020,cosco2021,weiss2021,pedernales2021} which add to other tests based on superpositions of source masses \cite{lindner2005,bahrami2015,carlesso2017,carlesso2019,haine2021}.	Nevertheless, with the current state of the art these proposals are still extremely challenging to realize in practice \cite{schmole2016, pedernales2020a}.
	
	In their recent work \cite{carney2021b}, Carney, M\"uller and Taylor propose an interesting interferometric scheme for testing the ability of the gravitational interaction to act as a quantum channel under what appear to be significantly reduced experimental constraints. Notably, their proposal makes use of a light test mass in a double-well potential that is gravitationally interacting with a very heavy source mass which, however, does not need to be prepared in a pure quantum state, thus enabling the use of even larger and more massive particles. The central claim of \cite{carney2021b} is that under very reasonable assumptions a model of gravity, in which the interaction is mediated by a classical channel, can not produce the collapse and revival dynamics in the interferometric contrast of the test mass generated by a quantum gravitational interaction. Even more remarkably, the signature of the quantum interaction is enhanced by a finite temperature of the heavy source mass, facilitating the discrimination of the two cases.
	
	In this work, we explicitly construct a local operations and classical communications (LOCC) channel between a harmonic oscillator and a particle in a double-well potential, that is fully compatible with the conditions of the proof in \cite{carney2021b} and reproduces the collapse-and-revival dynamics in the interferometric signal. This allows us to identify the error in the proof and leads us to the conclusion that the protocol presented in \cite{carney2021b} does not constitute a sufficient test to determine the nature of the gravitational interaction. We then proceed to discuss certain tests of LOCC models and analyse the consequences of such LOCC models for the physics of black holes.

	\section{Revivals Due to a Coherent Interaction}
	\label{sec:original-proposal}
\begin{figure}[h]
		\centering
		\includegraphics[width=0.75\columnwidth]{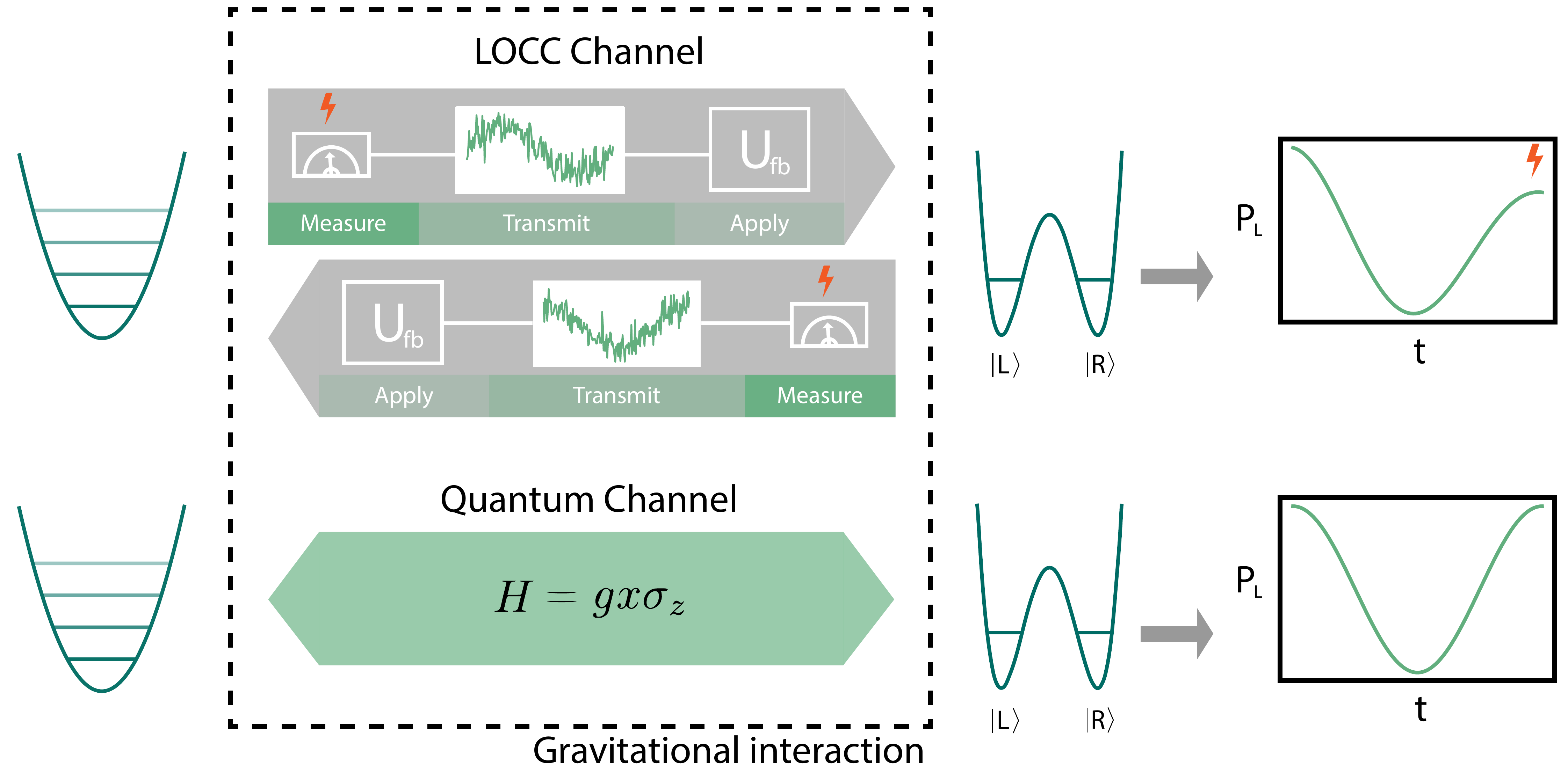}
		\caption{A particle trapped in a harmonic potential (\textbf{left}) and initialized in a thermal state interacts gravitationally with a lighter particle in a double-well potential (\textbf{right}). If the gravitational interaction is inherently classical it must admit an effective description as a local operations and classical communications (LOCC) channel (upper half) between the two quantum systems. This in turn necessitates the conversion of quantum information into classical information via a measurement leading to back-action and therefore heating of the quantum systems (lightning bolt). This decoherence can be detected in an interferometric measurement on the lighter mass (see main text for details). Contrary to this case, if the gravitational interaction can be modeled as a quantum channel no excess decoherence is expected in the interferometric signal (lower part).}
		\label{fig:overview}
	\end{figure}

	The system studied in \cite{carney2021b} consists of a large particle (A) trapped in a harmonic potential and an atom (B) trapped in a double-well potential as shown in Figure~\ref{fig:overview}. The double-well potential localizes the atom to two positions allowing an effective description of the spatial degree of freedom as a two-level system that we call a two-level test mass (TLTM). We, therefore, set the position operator of the atom to the Pauli z matrix $x_b = l\sigma_z$ with the eigenstates $\ket{L}$ and $\ket{R}$, which correspond to the atom occupying the left and right well, respectively and the factor $l$ denotes the separation distance between the two wells. The two systems interact gravitationally via a linearized Newtonian potential leading to the following Hamiltonian
	\begin{equation}
		H = \hbar \omega a^\dag a + \hbar g(a + a^\dag) \sigma_z,
		\label{eq:hamiltonian}
	\end{equation}
	with $a^\dag$ and $a$ denoting the creation and annihilation operators of the harmonic oscillator. For brevity, the energy splitting term for the TLTM has been omitted as one can always transform into a rotating frame without affecting the interaction term. Up to a global phase, the evolution operator for  Hamiltonian~(\ref{eq:hamiltonian}) can be written as \cite{carney2021b}
	\begin{equation}
		U(t) = D^\dag(\sigma_z \lambda) e^{-i \omega a^\dag a t} D(\sigma_z \lambda),
	\end{equation}
	with $\lambda = g/\omega$ and $D(\sigma_z \lambda)=\exp[(\lambda a^\dag - \lambda^* a) \sigma_z]$ denoting the standard displacement operator.
	
	\onecolumngrid
	
	\twocolumngrid
	
	The aim of the protocol proposed in \cite{carney2021b} is to create a witness for the entangling character of the gravitational interaction. For that matter they propose to initialize the harmonic oscillator in its ground state $\ket{0}_A$ and the TLTM in a superposition of the $\sigma_z$ eigenstates $\ket{L}_B$ and $\ket{R}_B$. This leads to the following evolution
	\begin{equation}
		\begin{aligned}
		\ket{\psi(t)} &= U(t) \left(\ket{0}_A \otimes \frac{1}{\sqrt{2}} \left( \ket{L}_B + \ket{R}_B \right)\right) \\&= \frac{1}{\sqrt{2}}\left( \ket{\delta}_A\ket{L}_B + \ket{-\delta}_A\ket{R}_B \right),
	\end{aligned}
	\end{equation}
	where $\ket{\pm \delta}$ denote coherent states with amplitudes $\delta~=~\pm \lambda \left( e^{-i \omega t} - 1 \right)$. The interaction term leads to opposite displacements of the oscillator depending on the TLTM state, thus building-up entanglement. In a final step, a $\pi/2$-pulse is applied to the TLTM mapping the signal to a population difference in the z basis and leading to the final state
	\begin{multline}
		\ket{\psi_f} = \frac{1}{2} \left( \ket{\delta}_A + \ket{-\delta}_A \right) \otimes \ket{L}_B \\+ \frac{1}{2} \left( \ket{\delta}_A - \ket{-\delta}_A \right) \otimes \ket{R}_B.
	\end{multline}
	The population of the $\ket{L}_B$ TLTM state is subsequently measured leading to the following signal
	\begin{equation}
		P_B(L) = \frac{1}{2} \left( 1 + e^{-8\lambda^2 \sin^2(\omega t/2)}\right).
	\end{equation} 
	Furthermore, an analogous calculation with the oscillator in a thermal initial state, with mean occupation number $\bar n$, yields
	\begin{equation}
		\tilde{P}_{B}(L) = \frac{1}{2} \left( 1 + e^{-8\lambda^2 (\bar{n} + 1) \sin^2(\omega t/2)}\right).
		\label{eq:gs-population-th}
	\end{equation}
	We observe that this function is oscillating with the frequency of the oscillator and is unity for multiples of $t=2 \pi/ \omega$. At these times the oscillator returns to its initial state and the entanglement between the two systems vanishes which can be seen in the form of Equation~(\ref{eq:gs-population-th})  given above. This property is not only preserved for thermal initial states of the oscillator but the contrast in the collapse and revival is even enhanced, significantly relaxing the experimental conditions required to observe it.
	
	The central result of \cite{carney2021b} is that the oscillatory signal produced by the coherent coupling term in Equation~(\ref{eq:hamiltonian}) is a signature of a quantum interaction that can not be reproduced by a classical one, i.e., by a separable quantum channel. In the next section, we give an explicit counterexample to this claim.
	
	\section{LOCC Model}
	In this section, we present a model where the position of the harmonic oscillator (A) is continuously measured via a homodyne measurement and the results are applied to the TLTM (B) via feedback. The phase of the TLTM is simultaneously measured via another homodyne measurement and fed back to the oscillator creating an interaction that can transmit classical information between both systems but is incapable of creating entanglement.
	
	The conditional state of a harmonic oscillator subject to a homodyne measurement is governed by the following stochastic master equation (SME) \cite{wiseman2009}
	\begin{multline}
		\textrm{d}\rho_{A,W} = -\frac{i}{\hbar} [H_0, \rho_{A,W}] \textrm{d}t \\ + \alpha^2 \mathcal{D}[x] \rho_{A,W} \textrm{d}t + \alpha \textrm{d}W_A \mathcal{H}[x] \rho_{A,W},
		\label{eq:homodyne-a}
	\end{multline}
	with $H_0$ denoting the free evolution Hamiltonian, $\mathcal{D}[x]$ the standard Lindbald dissipator and $\mathcal{H}[x]$
	\begin{equation}
		\mathcal{H}[ x ]\rho = x \rho + \rho x^\dag	 - \braket{x + x^\dag} \rho.
	\end{equation}
	$\textrm{d}W_A$ denotes a Gaussian noise term with zero mean $\textrm{E}[\textrm{d}W_A] = 0$ and variance $\textrm{E}[\textrm{d}W_A^2] = \textrm{d}t$. The position of the oscillator is encoded in the homodyne current produced by the measurement
	\begin{equation}
		J_A(t) = 2 \alpha \langle x \rangle + \textrm{d}W_A(t)/{\textrm{d}t}.
	\end{equation}
	
	Before we discuss the feedback we derive the SME for a TLTM subjected to a homodyne measurement in greater detail. The Kraus operators describing the continuous measurement of the TLTM in the z-basis are proportional to the $\ketbra{1}$ projector and given~by
	\begin{align}
		K_1 &= \frac{\beta}{2} \left( \sigma_z + \idop \right) \sqrt{dt}, \\
		K_0 &= 1 - \frac{\beta^2}{4} \left( \sigma_z + \idop \right) dt.
	\end{align}
	However, we study a model where the interaction induces a phase on the TLTM without changing the population in the z-basis, hence we choose to instead measure that phase by projecting on the $\ket{\pm}$ eigenstates of the $\sigma_x$ operator. The corresponding Kraus operators are given by the linear combinations of operators for the z-basis measurement \cite{gross2018}
	\begin{align}
		K_\pm = \frac{1}{\sqrt{2}} \left( 1 \pm \beta \left( \sigma_z + \idop \right) \sqrt{dt} - \frac{\beta^2}{4} \left( \sigma_z + \idop \right) dt \right).
	\end{align}
	The corresponding POVM is 
	\begin{equation}
		E_\pm = K_\pm^\dag K_\pm = \frac{1}{2} \left( 1 \pm \beta (\sigma_z + \idop) \sqrt{dt} \right).
	\end{equation}
	This leads to a SME that depends on the measurement outcomes
	\begin{multline}
		\textrm{d} \rho_\pm = \\ \left( \pm \beta \sqrt{dt} - \beta^2 \braket{\sigma_z + \idop} \textrm{d}t \right) \mathcal{H}\left[ \frac{\beta}{2} \left( \sigma_z + \idop \right)\right] \rho \\ + \mathcal{D}\left[ \frac{\beta}{2} \left( \sigma_z +  \idop \right)\right] \rho \textrm{d}t.
		\label{eq:homodyne-precursor-TLTM}
	\end{multline}
	Note that the dissipator is equivalent to the dephasing term
	\begin{equation}
		\mathcal{D}[\sigma_z]\rho = \frac{\beta^2}{4} \left( \sigma_z \rho \sigma_z - \rho\right).
	\end{equation}
	The dependence on the measurement result only enters the first term $dM(t) = \pm \sqrt{dt}$ in Equation~(\ref{eq:homodyne-precursor-TLTM}). It is easy to show that this term obeys Gaussian statistics 
	\begin{equation}
		\begin{aligned}
		\textrm{E}[dM(t)] &= \beta \braket{\sigma_z +\idop} \textrm{d}t,\\ \textrm{E}[dM(t)^2] &= dt.
		\end{aligned}
	\end{equation}
	Furthermore, we note that the mean corresponds to the second term in Equation~(\ref{eq:homodyne-precursor-TLTM}) which enables us to rewrite Equation~(\ref{eq:homodyne-precursor-TLTM}) in terms of a zero-mean Gaussian noise term that we label with $dW_{B}(t)$
	\begin{equation}
		\textrm{d} \rho_{B,W} = \frac{\beta^2}{4} \mathcal{D}\left[\sigma_z \right] \rho_{B,W}\textrm{d}t + \frac{\beta}{2} dW_B(t) \mathcal{H}\left[ \sigma_z \right] \rho_{B, W}.
		\label{eq:homodyne-b}
	\end{equation}
	{Note} that we have omitted the Hamiltonian term for the TLTM as it is irrelevant for our analysis. Finally, we identify the homodyne current as
	\begin{equation}
		J_B(t) =  \beta  \braket{\sigma_z + \idop} + \textrm{d}W_B(t)/{\textrm{d}t}.
	\end{equation}
	Having obtained a classical stream of information for each system in the form of two homodyne currents we now set out to couple them via feedback. It is important to note that the measurement results encoded in $J_A(t)$ and $J_B(t)$ are not experimentally accessible as they model an internal process of the gravitational interaction. To determine the experimentally observable quantities we have to obtain the unconditional state of the system, i.e., we have to average over the noise terms $\textrm{d}W_{A,B}(t)$. 
	
\begin{figure}[t]
		%  \centering
		\includegraphics[width=\columnwidth]{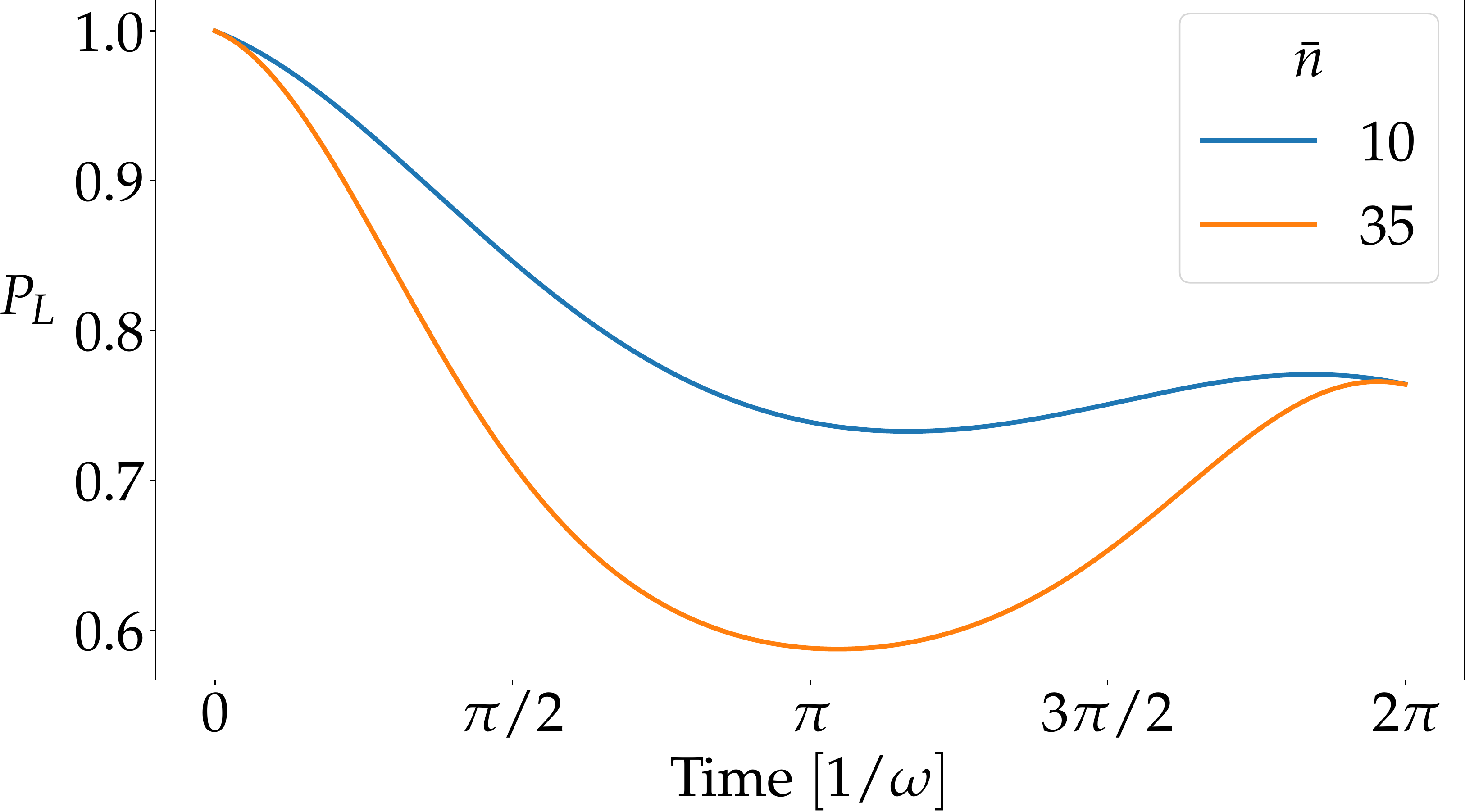}
		\caption{Revivals in the population of the $\ket{L}$ state of the TLTM for different initial thermal states of the oscillator for the model in Equation~(\ref{eq:unconditional-me}) and the protocol described in Section \ref{sec:original-proposal}. For the simulations presented here, we chose $2\alpha\beta = 2 \alpha^2 = \beta^2 /2 = 0.05\ \omega$.}
		\label{fig:revivals}
	\end{figure}

	We defer the details of this calculation to Appendix~\ref{sec:appendix-feedback-eq}. Here, we note that our LOCC channel consists of two parts, each containing a measurement on one system and feedback on the other. Each part yields the following contribution to the master equation
	\begin{equation}
		[\dot{\rho}]_{i\rightarrow j} = -i \frac{1}{2}[M^\dag F + M F, \rho] + \mathcal{D}[ M -i F]\rho.
		\label{eq:uni-directional-locc}
	\end{equation}
	where the measurement operator $M$ acts on system $i$ and the feedback operator $F$ acts on system $j$. Combining these two parts yields a master equation in Lindblad form as the terms proportional to $F\rho M$ cancel
	\begin{multline}
		\dot{\rho} = -\frac{i}{\hbar} [H_0, \rho] - i \alpha \beta [ x, \rho] - 2 i \alpha \beta [ x \sigma_z, \rho] \\ + 2 \alpha^2 \mathcal{D}[x] \rho + \frac{\beta^2}{2} \mathcal{D}[\sigma_z] \rho.
		\label{eq:unconditional-me}
	\end{multline}
	We recognize the linear coupling term as it also appears in the expansion of the gravitational potential \cite{pedernales2021}, showing that this model reproduces the phenomenology of the gravitational interaction. Furthermore, we note that it is this term that leads to the appearance of revivals in the protocol of \cite{carney2021b} which we demonstrate with a numerical simulation in Figure~\ref{fig:revivals}. The difference of this model to one with a purely quantum interaction is the appearance of decoherence terms in very much the same way as in the LOCC coupling of two harmonic oscillators \cite{kafri2014}. This dynamical equation by construction corresponds to a classical interaction channel that can not generate entanglement. Yet, it contains a term that, according to the proof of \cite{carney2021b}, is incompatible with an LOCC channel. In the next section, we explore the origin of this discrepancy.
	
	\section{Product Form Kraus Representation for the LOCC Model}
	Given a master equation in diagonal Lindblad form 
	\begin{equation}
		\dot{\rho} = -\frac{i}{\hbar} [H, \rho] + \sum_{i=1}^N \left( E_i \rho E_i^\dag - \{ E_i^\dag E_i, \rho \} \right),
	\end{equation}
	one can immediately determine the Kraus representation to be \cite{wiseman2009}
	\begin{equation}
		\begin{aligned}
		L_0 &= \idop - \frac{i}{\hbar} H \textrm{d}t - \frac{1}{2} \sum_{i=1}^N E_i^\dag E_i \textrm{d}t \\ L_i &= E_i \sqrt{\textrm{d}t}.
		\end{aligned}
		\label{eq:lindbald-kraus-decomposition}
	\end{equation}
	This is the same form as chosen in \cite{carney2021b}. For Equation~(\ref{eq:unconditional-me}) we have $E_1=x$, $E_2=\sigma_z$ and a Hamiltonian with an interaction term $H = H_0 + 2 \alpha \beta x \sigma_z$. It is this term that makes this channel seemingly non-separable by the arguments presented in \cite{carney2021b}. However, the Kraus operators are not unique but can be subjected to a unitary transformation without changing the underlying operation. Therefore, a statement about the separability of a channel can not be made based on a specific Kraus decomposition, rather it has to be shown that no separable decomposition exists.
	
	To find a separable Kraus representation for the LOCC model presented in the previous section we start with Equation~(\ref{eq:uni-directional-locc}). While it models only a uni-directional interaction the inverse channel is completely symmetric, hence the form of its Kraus operators will be identical but with measurement and feedback operators interchanged. These two sets of operators can then be readily combined to yield a separable representation for Equation~(\ref{eq:unconditional-me}). Note that Equation~(\ref{eq:uni-directional-locc}) does not contain the $H_0$ term present in Equation~(\ref{eq:unconditional-me}) because it is not relevant for our analysis and can be easily added to the final form of the Kraus operators in the same way as the Hamiltonian term in Equation~(\ref{eq:lindbald-kraus-decomposition}). By the above procedure, we read off the Kraus operators for Equation~(\ref{eq:uni-directional-locc}) 
	\begin{equation}
		\begin{aligned}
			L_0 &=  \idop - i M F \textrm{d}t - \frac{1}{2} ( M^2 + F^2) \textrm{d}t, \\
			L_1 &= ( M - i F)\sqrt{\textrm{d}t}.
		\end{aligned}
	\end{equation}
	Here we assumed $M = M^\dag$. $L_1$ is not of product form as it is a sum of two product form operators that are both proportional to $\sqrt{\textrm{d}t}$, furthermore the term $i M F \textrm{d}t$ also prevents us from writing $L_0$ in product form. We now apply the unitary transformation 
	\begin{align}
		\begin{pmatrix}
			\tilde{L}_0 \\
			\tilde{L}_1
		\end{pmatrix} =
		\frac{1}{\sqrt{2}}
		\begin{pmatrix}
			1 & -1 \\
			1 & 1
		\end{pmatrix}
		\begin{pmatrix}
			L_0 \\
			L_1
		\end{pmatrix},
		\label{eq:transformation}
	\end{align}
	obtaining the following Kraus operators
	\begin{equation}
		\begin{aligned}
			\tilde{L}_0 =  \frac{1}{\sqrt{2}} \bigg( \idop - ( M - i F)&\sqrt{\textrm{d}t} \\ - i M F \textrm{d}t & - \frac{1}{2} ( M^2 + F^2) \textrm{d}t \bigg), \\
			\tilde{L}_1 = \frac{1}{\sqrt{2}} \bigg( \idop + ( M - i F)&\sqrt{\textrm{d}t} \\- i M F \textrm{d}t &- \frac{1}{2} ( M^2 + F^2) \textrm{d}t \bigg).
		\end{aligned}
	\end{equation}
	These can now be written in product form
	\begin{align}
	\begin{aligned}
		\tilde{L}_0 = \frac{1}{\sqrt{2}} &\left( \idop - M \sqrt{\textrm{d}t} - \frac{1}{2}M^2 \textrm{d}t \right)\\ \otimes &\left( \idop + iF \sqrt{\textrm{d}t} - \frac{1}{2} F^2 \textrm{d}t \right),
	\end{aligned}\\
	\begin{aligned}
		\tilde{L}_1 = \frac{1}{\sqrt{2}} &\left( \idop +  M \sqrt{\textrm{d}t} - \frac{1}{2}M^2 \textrm{d}t \right) \\  \otimes & \left( \idop - i F \sqrt{\textrm{d}t} - \frac{ 1 }{2} F^2 \textrm{d}t \right).
	\end{aligned}
	\end{align}
	To obtain the action of both parts of the LOCC channel we need a second set of Kraus operators $\{\tilde{L}'_0, \tilde{L}'_1\}$ with operators $M$ and $F$ interchanged. Both sets must be combined in the following way
	\begin{equation}
		K_{ij} = \tilde{L}'_i \tilde{L}_j,
	\end{equation}
	which produces a total of four Kraus operators that by construction remain separable.
	
	Comparing the form of the Kraus operators obtained here and those used in the proof of \cite{carney2021b} we realize a crucial difference: the presence of terms proportional to $\sqrt{\textrm{d}t}$ in all Kraus operators. It is the restriction to linear terms in $\textrm{d}t$ for the $L_0$ operator that leads Carney et al. to the erroneous conclusion that a separable channel can not produce an interaction term as it appears in Equation~(\ref{eq:unconditional-me}). We, therefore, conclude that the absence of revivals can not serve as a conclusive test of the quantum nature of gravity. However, the LOCC model does introduce additional decoherence which can in principle be experimentally measured.
	
	\section{Experimental Tests of Gravitational Decoherence}
	The Newtonian interaction potential merely constrains the interaction term in \linebreak \mbox{Equation~(\ref{eq:unconditional-me})} to \cite{pedernales2021}
	\begin{equation}
		g = 2 \alpha \beta = \frac{G m_a m_b}{ \hbar d^3} l x_0,
		\label{eq:newtonian-constraint}
	\end{equation}
	where we have denoted the separation distance of the two wells that make up the TLTM with $l$ and $x_0 = \sqrt{\hbar/(2 \omega m_a)}$ denoting the size of the ground state wavefunction for the harmonic oscillator. By introducing $x_0$ here, we assume that the $x$ operator that appears in Equation~(\ref{eq:unconditional-me}) has the dimensionless form $x = a+a^\dag$. This leaves substantial freedom in the distribution of the additional noise induced by the LOCC channel. It was previously noted in \cite{kafri2014} that the total amount of noise is minimized for a symmetric splitting
	\begin{equation}
		\gamma_s = 2\alpha^2 = \frac{\beta^2}{2} = \frac{G m_a m_b }{ 2 \hbar d^3} l x_0 .
		\label{symmetric}
	\end{equation}
	{Using} this choice, we estimate the heating rate of an atom with the parameters stated \mbox{in \cite{carney2021b}}, i.e., $m_a=1$~mg, $m_b=133$~amu, $l \approx d = 1$~mm, to be $\gamma_s \approx 2\pi \times 10^{-15}$~Hz. In an experiment with two levitated silica spheres (density $2.65$~g/cm$^3$) with radii $R=1~\mu$m and $r=100$~nm at a distance $d=2R$, with the smaller particle in a double-well potential with separation $l=1~\mu$m and the larger particle in a harmonic potential with $\omega=2\pi \times 10^2$ Hz, we would expect a heating rate of $\gamma_s \approx 2\pi \times 10^{-8}$~Hz. In both cases the heating rates are many orders of magnitude lower than what is reported in current experiments \cite{delic2020, gieseler2020}, making their detection extremely challenging. It was argued in \cite{altamirano2018} that the case of symmetric splitting Equation~(\ref{symmetric}) can be ruled out by the experiments of \cite{kovachy2015}, a case that has however faced some debate \cite{stamper-kurn2016, kovachy2016}.
	
	Another appealing choice is one for which the decoherence rate of a particle will only depend on its own mass and a distance, $d$, to other gravitating masses, that is
	\begin{equation}
		\gamma_a = 2 \alpha^2 = \frac{GM^2}{2\hbar d^3} x_0^2.
		\label{eq:asymmetric-noise}
	\end{equation}
	{Such} a choice means that the noise is virtually impossible to detect with light particles such as atoms. However, it increases the noise on the larger mass, s.t. for the parameters in~\cite{carney2021b} we obtain $\gamma_a \approx 2\pi \times 10^{-6}$~Hz. The detection of this heating rate can be facilitated by entangling the heavy mass with another sensor system, as it is proposed with the boosted protocol in \cite{carney2021b}, as well as in \cite{pedernales2021}, such a heating rate could potentially be more easily detected. We note that our proposal \cite{pedernales2021} would not only allow to measure this heating rate via the strongly coupled TLTM, but also provide a second test by allowing direct verification of entanglement generation. Finally, for this choice of noise coupling the heating rate on a superposition of $1~\mu$m of a silica sphere with a radius of 100~nm would also correspond to $\gamma_a \approx 2\pi \times 10^{-6}$~Hz.
	
	The choice in Equation~(\ref{eq:asymmetric-noise}) still depends on the distance to a particular second mass. A reasonable demand is that the heating rate should only depend on local quantities. A suitable choice would be to set $d$ to the particle radius, as this choice ensures that no entanglement is generated even at the closest approach of another body. For this choice the heating rate becomes proportional to the mass density. This of course breaks the relationship between the decoherence rates $\alpha^2$, $\beta^2$ and the coupling term $\alpha\beta$. However, this relationship is not as strict as is suggested in Equation~(\ref{eq:unconditional-me}). Arguably the result of the back-action inducing measurement can be used to set a classical potential strength which then falls off via the typical inverse-square law, i.e., the measurement result is broadcast to all other masses. For the analysis of this case we consider a neutron, trapped in a harmonic potential with frequency $\omega = 2 \pi \times 220$ Hz, as in the experiment of \cite{cronenberg2018}. Using its Compton wavelength for the value of $d$ in Equation~(\ref{eq:asymmetric-noise}) we obtain a value $\gamma_n = 2\pi \times 2.9$ kHz. In \cite{schimmoller2021} a thorough comparison between the experimental data of \cite{cronenberg2018} and a model with the same heating term as the one considered here was performed. The conclusion therein is that heating rates larger than $\gamma_n \approx 1$ Hz are not compatible with the experimental data in \cite{cronenberg2018}. Therefore our estimate is not compatible with the experimental data; however, we have implicitly assumed that the form of the Newtonian potential is valid down to femtometer scale, a regime which is so far experimentally unexplored.
	
	The work in \cite{schimmoller2021} further shows that the detection of an excess heating rate on only one of the two systems is not sufficient to rule out a quantum character of the gravitational interaction, because such a heating rate also appears if gravity is modeled as an entropic force \cite{verlinde2011}.
	
	\section{A Remark on Classical Gravity and Black Hole Radiation}
	In this section, we would like to explore further the choice of a heating rate that is only dependent on the properties of the object itself. To that end we consider the most extreme case possible, namely that of a Schwarzschild black hole. With the extreme choice $d = r_S = 2GM/c^2$, i.e., the Schwarzschild radius, the heating rate of the black hole induced by the LOCC channel is given by
	\begin{equation}
		\frac{d}{dt}\langle{\frac{p^2}{2M}}\rangle = \frac{GM^2\hbar}{M r_S^3}
		= \frac{\hbar c^6}{8 M^2 G^2}.
		\label{LOCCHeating}
	\end{equation}
	{In} order to achieve an equilibrium within a theory of classical gravity the black hole will need to lose this amount of energy via black body radiation as seen by an external observer. Remarkably, the radiative power predicted by Equation~(\ref{LOCCHeating}) is in fact proportional to the power of the Hawking radiation expected from such a black hole under the assumption of pure photon emission from the surface at the Schwarzschild radius, $P = \hbar c^6 / (15360 \pi M^2 G^2)$. Naturally, we do not expect a perfect match, as Equation~(\ref{LOCCHeating}) does not account for the fact that the assumption of Newtonian gravity is certainly inaccurate close to the black hole horizon. The appearance of a certain multiple of the Schwarzschild radius is therefore expected.
	
	Note that the consequences of the predicted radiation appear to be very different under Hawking's mechanism and under the LOCC gravity model. In the former, the emitted radiation leads to the eventual evaporation of the black hole bringing with it the hotly debated information paradox. In the latter, the black hole would be eternal as the emitted radiation would merely balance the heating generated through the required noise in an LOCC model of gravity and would thus violate energy conservation. Neither seems very palatable to physicists. In any case, it is intriguing that the LOCC-mediated gravity leads black holes to emit thermal radiation with similar properties, integrated power and thermality, to that predicted of Hawking radiation.
	
	\section{Conclusions} It should be noted that independent of our work the proposal of Carney, M{\"u}ller and Taylor \cite{carney2021b} has, to the best of our knowledge, been criticised in two recent independent works \cite{hosten2021, ma2021} that highlight that the signature claimed to be unique to an entangling channel can be reproduced by semi-classical models. However, unlike our work, {refs.}~\cite{hosten2021, ma2021} deviate significantly from the assumptions of theorem 1 in \cite{carney2021b} and are thus not conclusive. {Ref.}~\cite{hosten2021} treats the trajectory of the massive particle classically. This omits any effect of measurement back-action involved in mapping a quantum mechanical observable to a classical one as is required in the attempt of realizing a gravitational interaction as a classically mediated force. The resulting additional noise could potentially destroy the revival dynamics, therefore {ref.}~\cite{hosten2021} does not invalidate theorem 1 of \cite{carney2021b}. {Ref.}~\cite{ma2021} provides an example which, as the authors stress, violates the assumption of time-translation invariance that enters the proof of theorem 1 in \cite{carney2021b} in an essential manner. Hence, neither of these two works conclusively falsifies the claims of \cite{carney2021b}.
	
	In contrast, we have demonstrated that while fulfilling all assumptions of the proof in \cite{carney2021b} an LOCC model can reproduce revivals in the interferometric signal contradicting theorem 1 of \cite{carney2021b}. Furthermore, we discussed the heating rate expected from such an LOCC model and highlighted its connection to the Hawking radiation.
	
	\begin{acknowledgments}
	This research was funded by the ERC Synergy grant HyperQ (Grant No. 856432), the EU project AsteriQs (Grant No. 820394), the BMBF project DiaPol (13GW 0281C) and the state of Baden-W\"urttemberg through bwHPC, the German Research Foundation (DFG) through Grant No. INST 40/467-1 FUGG. The results in this work were discussed with Carney, M{\"u}ller and Taylor. Their conclusions were posted on the same day as this work on the arXiv \cite{carney2021d}.
	\end{acknowledgments}
	
	\appendix
	\section{Derivation of the Feedback Equation}
	\label{sec:appendix-feedback-eq}
	In this appendix we derive Equation~(\ref{eq:uni-directional-locc}) of the main text. The total SME for the combined state $\rho$ of the harmonic oscillator and TLTM under their respective homodyne measurements evolves according to (the $H_0$ term acting on the oscillator is omitted)
	\begin{multline}
			\textrm{d}\rho(t) = \alpha^2 \mathcal{D}[x] \rho(t) \textrm{d}t + \alpha \textrm{d}W_A \mathcal{H}[x]\rho(t) \\ + \frac{\beta^2}{4} \mathcal{D}\left[\sigma_z \right] \rho(t) \textrm{d}t + \frac{\beta}{2} dW_B(t) \mathcal{H}\left[ \sigma_z \right] \rho(t).
	\end{multline}
	{We} assume that the feedback is performed via a unitary operation of the form
	\begin{align}
		[\dot{\rho}_A(t)]_{fb} = -i J_B(t) [F_A, \rho_A(t)], \\
		[\dot{\rho}_B(t)]_{fb} = -i J_A(t) [F_B, \rho_B(t)].
	\end{align}
	where $F_A = \alpha x$ and $F_B = \beta \sigma_z /2$ and the homodyne currents are given by
	\begin{align}
		J_A(t) &= 2 \alpha \langle x \rangle + \textrm{d}W_A(t)/{\textrm{d}t}, \\
		J_B(t) &=  \beta  \braket{\sigma_z + \idop} + \textrm{d}W_B(t)/{\textrm{d}t}.
	\end{align}
	{It} is shown in \cite{wiseman2009} that the feedback superoperator takes the following form
	\begin{multline}
		\mathcal{F}(t)\rho(t) = \exp\Big( -i J_A(t) [F_B, \rho(t)] \textrm{d}t \\-i J_B(t) [F_A, \rho(t)]\textrm{d}t \Big),
	\end{multline}
	and enters the SME in a multiplicative way
	\begin{multline}
			\textrm{d}\rho(t) = \mathcal{F}(t) ( \alpha^2 \mathcal{D}[x] \rho(t) \textrm{d}t + \alpha \textrm{d}W_A \mathcal{H}[x]\rho(t) \\
			+ \frac{\beta^2}{4} \mathcal{D}\left[\sigma_z \right] \rho(t) \textrm{d}t + \frac{\beta}{2} dW_B(t) \mathcal{H}\left[ \sigma_z \right] \rho(t)).
			\label{eq:full-sme}
	\end{multline}
	{As} only terms up to linear order in d$t$ are relevant we can expand $\mathcal{F}(t)$ while keeping in mind the properties of the Gaussian noise terms. Furthermore, as already noted in the main text, we are interested in the unconditional evolution of $\rho(t)$, i.e., we need to perform the average over $\textrm{d}W_{A,B}(t)$. This means that terms linear in $\textrm{d}W_{A,B}(t)$ or of mixed form $\textrm{d}W_{A}(t) \textrm{d}W_{B}(t)$, will average to zero and can therefore already be neglected. Therefore, we~obtain 
	\begin{equation}
		\begin{aligned}
			\mathcal{F}(t)\rho(t) = \idop &- i \alpha \beta \braket{x} [x, \rho(t)] \textrm{d}t - i \frac{\beta}{2} [\sigma_z, \rho(t)] \textrm{d}W_A(t) \\
			&- \frac{\beta^2}{8} [\sigma_z, [\sigma_z, \rho(t)]] \textrm{d}t - i \alpha \beta \braket{\sigma_z} [x, \rho(t)] \textrm{d}t  \\
			&-i \alpha \beta [x, \rho(t)] \textrm{d}t - i \alpha \beta [x, \rho(t)] \textrm{d}W_B(t) \\
			&- \frac{\alpha^2}{2} [x, [x, \rho(t)]] \textrm{d}t,
		\end{aligned}
	\end{equation}
	where we have used $\textrm{E}[\textrm{d}W_{A,B}(t)^2] = \textrm{d}t$ which led to the double commutator terms that translate into the dissipator terms in the final equation. Inserting this form into \mbox{Equation~(\ref{eq:full-sme})} and averaging over the noise terms produces Equation~(\ref{eq:unconditional-me}) in the main text. That equation is obtained by passing via the intermediate form
	\begin{equation}
		\begin{aligned}
			\dot{\rho}(t) &= i \alpha \beta [x, \rho(t)] - i \frac{\alpha \beta}{2} [x, \sigma_z \rho(t) + \rho(t) \sigma_z] + \alpha^2 \mathcal{D}[x]\rho(t) \\
			&+\frac{\beta^2}{4}\mathcal{D}[\sigma_z]\rho(t) - i\frac{\alpha \beta}{2} [\sigma_z, x\rho(t) + \rho(t) x] + \alpha^2 \mathcal{D}[x]\rho(t) \\
			&+ \frac{\beta^2}{4}\mathcal{D}[\sigma_z]\rho(t),
		\end{aligned}
	\end{equation}
	which itself consists of two parts each of which contributes a one directional channel where a measurement is performed on one system and the feedback operations is applied to the other. Each of these sub-parts can be further rearranged to yield the diagonal Lindblad form in Equation~(\ref{eq:uni-directional-locc}).
	
	\bibliographystyle{apsrev4-2}

\begin{thebibliography}{99}
			
			\bibitem[Rickles and DeWitt (2011)]{Feynman57}
			Rickles, D.; DeWitt, C. M.
			\newblock {\em The Role of Gravitation in Physics: Report from the 1957 Chapel Hill Conference;}
			\newblock {Max-Planck-Gesellschaft zur F\"orderung der Wissenschaften}: Berlin, Germany, 2011; p 250ff.			
			
			\bibitem[Karolyhazy(1966)]{karolyhazy1966}
			Karolyhazy, F.
			\newblock Gravitation and Quantum Mechanics of Macroscopic Objects.
			\newblock {\em Nuovo C. A (1965--1970)} {\bf 1966}, {\em 42},~390--402.
			\newblock
			[\href{http://doi.org/10.1007/BF02717926}{CrossRef}]
			
			\bibitem[Penrose(1996)]{penrose1996}
			Penrose, R.
			\newblock On {{Gravity}}'s Role in {{Quantum State Reduction}}.
			\newblock {\em Gen. Relat. Gravit.} {\bf 1996}, {\em 28},~581--600.
			\newblock
			[\href{http://dx.doi.org/10.1007/BF02105068}{CrossRef}]
			
			\bibitem[Kafri and Taylor(2013)]{kafri2013}
			Kafri, D.; Taylor, J.M.
			\newblock A Noise Inequality for Classical Forces.
			\newblock {\em arXiv} {\bf 2013}, \href{https://doi.org/10.48550/arXiv.1311.4558}{arXiv:1311.4558}.
			
			\bibitem[{Gonzalez-Ballestero} \em{et~al.}(2021){Gonzalez-Ballestero},
			Aspelmeyer, Novotny, Quidant, and {Romero-Isart}]{gonzalez-ballestero2021}
			{Gonzalez-Ballestero}, C.; Aspelmeyer, M.; Novotny, L.; Quidant, R.;
			{Romero-Isart}, O.
			\newblock Levitodynamics: {{Levitation}} and Control of Microscopic Objects in
			Vacuum.
			\newblock {\em Science} {\bf 2021}, {\em 374},~eabg3027.
			\newblock
			[\href{http://dx.doi.org/10.1126/science.abg3027}{CrossRef}]
			
			\bibitem[Aspelmeyer \em{et~al.}(2014)Aspelmeyer, Kippenberg, and
			Marquardt]{Aspelmeyer2014}
			Aspelmeyer, M.; Kippenberg, T.J.; Marquardt, F.
			\newblock Cavity Optomechanics.
			\newblock {\em Rev. Mod. Phys.} {\bf 2014}, {\em 86},~1391--1452.
			\newblock
			[\href{http://dx.doi.org/10.1103/RevModPhys.86.1391}{CrossRef}]
			
			\bibitem[Deli{\'c} \em{et~al.}(2020)Deli{\'c}, Reisenbauer, Dare, Grass,
			Vuleti{\'c}, Kiesel, and Aspelmeyer]{delic2020}
			Deli{\'c}, U.; Reisenbauer, M.; Dare, K.; Grass, D.; Vuleti{\'c}, V.; Kiesel,
			N.; Aspelmeyer, M.
			\newblock Cooling of a Levitated Nanoparticle to the Motional Quantum Ground
			State.
			\newblock {\em Science} {\bf 2020}, {\em 367},~892--895.
			\newblock
			[\href{http://dx.doi.org/10.1126/science.aba3993}{CrossRef}] [\href{http://www.ncbi.nlm.nih.gov/pubmed/32001522}{PubMed}]
			
			\bibitem[Streltsov \em{et~al.}(2021)Streltsov, Pedernales, and
			Plenio]{streltsov2021}
			Streltsov, K.; Pedernales, J.S.; Plenio, M.B.
			\newblock Ground-{{State Cooling}} of {{Levitated Magnets}} in {{Low-Frequency
					Traps}}.
			\newblock {\em Phys. Rev. Lett.} {\bf 2021}, {\em 126},~193602.
			\newblock
			[\href{http://dx.doi.org/10.1103/PhysRevLett.126.193602}{CrossRef}]
			
			\bibitem[Pikovski \em{et~al.}(2012)Pikovski, Vanner, Aspelmeyer, Kim, and
			Brukner]{pikovski2012}
			Pikovski, I.; Vanner, M.R.; Aspelmeyer, M.; Kim, M.S.; Brukner, {\v C}.
			\newblock Probing {{Planck-scale}} Physics with Quantum Optics.
			\newblock {\em Nat. Phys.} {\bf 2012}, {\em 8},~393--397.
			\newblock
			[\href{http://dx.doi.org/10.1038/nphys2262}{CrossRef}]
			
			\bibitem[Kafri \em{et~al.}(2014)Kafri, Taylor, and Milburn]{kafri2014}
			Kafri, D.; Taylor, J.M.; Milburn, G.J.
			\newblock A Classical Channel Model for Gravitational Decoherence.
			\newblock {\em New J. Phys.} {\bf 2014}, {\em 16},~065020.
			\newblock
			[\href{http://dx.doi.org/10.1088/1367-2630/16/6/065020}{CrossRef}]
			
			\bibitem[Albrecht \em{et~al.}(2014)Albrecht, Retzker, and Plenio]{albrecht2014}
			Albrecht, A.; Retzker, A.; Plenio, M.B.
			\newblock Testing Quantum Gravity by Nanodiamond Interferometry with
			Nitrogen-Vacancy Centers.
			\newblock {\em Phys. Rev. A} {\bf 2014}, {\em 90},~033834.
			\newblock
			[\href{http://dx.doi.org/10.1103/PhysRevA.90.033834}{CrossRef}]
			
			\bibitem[Kumar and Plenio(2020)]{kumar2020}
			Kumar, S.P.; Plenio, M.B.
			\newblock On Quantum Gravity Tests with Composite Particles.
			\newblock {\em Nat. Commun.} {\bf 2020}, {\em 11},~3900.
			\newblock
			[\href{http://dx.doi.org/10.1038/s41467-020-17518-5}{CrossRef}] [\href{http://www.ncbi.nlm.nih.gov/pubmed/32764700}{PubMed}]
			
			\bibitem[Krisnanda \em{et~al.}(2017)Krisnanda, Zuppardo, Paternostro, and
			Paterek]{krisnanda2017}
			Krisnanda, T.; Zuppardo, M.; Paternostro, M.; Paterek, T.
			\newblock Revealing {{Nonclassicality}} of {{Inaccessible Objects}}.
			\newblock {\em Phys. Rev. Lett.} {\bf 2017}, {\em 119},~120402.
			\newblock
			[\href{http://dx.doi.org/10.1103/PhysRevLett.119.120402}{CrossRef}]
			
			\bibitem[Bose \em{et~al.}(2017)Bose, Mazumdar, Morley, Ulbricht, Toro{\v s},
			Paternostro, Geraci, Barker, Kim, and Milburn]{bose2017}
			Bose, S.; Mazumdar, A.; Morley, G.W.; Ulbricht, H.; Toro{\v s}, M.;
			Paternostro, M.; Geraci, A.A.; Barker, P.F.; Kim, M.S.; Milburn, G.
			\newblock Spin {{Entanglement Witness}} for {{Quantum Gravity}}.
			\newblock {\em Phys. Rev. Lett.} {\bf 2017}, {\em 119},~240401.
			\newblock
			[\href{http://dx.doi.org/10.1103/PhysRevLett.119.240401}{CrossRef}]
			
			\bibitem[Krisnanda \em{et~al.}(2020)Krisnanda, Tham, Paternostro, and
			Paterek]{krisnanda2020}
			Krisnanda, T.; Tham, G.Y.; Paternostro, M.; Paterek, T.
			\newblock Observable Quantum Entanglement Due to Gravity.
			\newblock {\em npj Quant. Inf.} {\bf 2020}, {\em 6},~12.
			\newblock
			[\href{http://dx.doi.org/10.1038/s41534-020-0243-y}{CrossRef}]
			
			\bibitem[Pedernales \em{et~al.}(2020)Pedernales, Morley, and
			Plenio]{pedernales2020}
			Pedernales, J.S.; Morley, G.W.; Plenio, M.B.
			\newblock Motional {{Dynamical Decoupling}} for {{Interferometry}} with
			{{Macroscopic Particles}}.
			\newblock {\em Phys. Rev. Lett.} {\bf 2020}, {\em 125},~023602.
			\newblock
			[\href{http://dx.doi.org/10.1103/PhysRevLett.125.023602}{CrossRef}] [\href{http://www.ncbi.nlm.nih.gov/pubmed/32701327}{PubMed}]
			
			\bibitem[Cosco \em{et~al.}(2021)Cosco, Pedernales, and Plenio]{cosco2021}
			Cosco, F.; Pedernales, J.S.; Plenio, M.B.
			\newblock Enhanced Force Sensitivity and Entanglement in Periodically Driven
			Optomechanics.
			\newblock {\em Phys. Rev. A} {\bf 2021}, {\em 103},~L061501.
			\newblock
			[\href{http://dx.doi.org/10.1103/PhysRevA.103.L061501}{CrossRef}]
			
			\bibitem[Weiss \em{et~al.}(2021)Weiss, {Roda-Llordes}, Torrontegui, Aspelmeyer,
			and {Romero-Isart}]{weiss2021}
			Weiss, T.; {Roda-Llordes}, M.; Torrontegui, E.; Aspelmeyer, M.; {Romero-Isart},
			O.
			\newblock Large {{Quantum Delocalization}} of a {{Levitated Nanoparticle Using
					Optimal Control}}: {{Applications}} for {{Force Sensing}} and {{Entangling}}
			via {{Weak Forces}}.
			\newblock {\em Phys. Rev. Lett.} {\bf 2021}, {\em 127},~023601.
			\newblock
			[\href{http://dx.doi.org/10.1103/PhysRevLett.127.023601}{CrossRef}]
			
			\bibitem[Pedernales \em{et~al.}(2021)Pedernales, Streltsov, and
			Plenio]{pedernales2021}
			Pedernales, J.S.; Streltsov, K.; Plenio, M.B.
			\newblock Enhancing {{Gravitational Interaction}} between {{Quantum Systems}}
			by a {{Massive Mediator}}.
			\newblock {\em Phys. Rev. Lett.} {\bf 2022}, {\em 128},~110401.
			\newblock
			[\href{https://doi.org/10.1103/PhysRevLett.128.110401}{CrossRef}]
			
			\bibitem[Lindner and Peres(2005)]{lindner2005}
			Lindner, N.H.; Peres, A.
			\newblock Testing Quantum Superpositions of the Gravitational Field with
			{{Bose-Einstein}} Condensates.
			\newblock {\em Phys. Rev. A} {\bf 2005}, {\em 71},~024101.
			\newblock
			[\href{http://dx.doi.org/10.1103/PhysRevA.71.024101}{CrossRef}]
			
			\bibitem[Bahrami \em{et~al.}(2015)Bahrami, Bassi, McMillen, Paternostro, and
			Ulbricht]{bahrami2015}
			Bahrami, M.; Bassi, A.; McMillen, S.; Paternostro, M.; Ulbricht, H.
			\newblock Is {{Gravity Quantum}}?
			\newblock {\em arXiv} {\bf 2015}, \href{https://doi.org/10.48550/arXiv.1507.05733}{arXiv:1507.05733}.
			
			\bibitem[Carlesso \em{et~al.}(2017)Carlesso, Paternostro, Ulbricht, and
			Bassi]{carlesso2017}
			Carlesso, M.; Paternostro, M.; Ulbricht, H.; Bassi, A.
			\newblock When {{Cavendish}} Meets {{Feynman}}: {{A}} Quantum Torsion Balance
			for Testing the Quantumness of Gravity.
			\newblock {\em arXiv} {\bf 2017}, \href{https://doi.org/10.48550/arXiv.1710.08695}{arXiv:1710.08695}.
			
			\bibitem[Carlesso \em{et~al.}(2019)Carlesso, Bassi, Paternostro, and
			Ulbricht]{carlesso2019}
			Carlesso, M.; Bassi, A.; Paternostro, M.; Ulbricht, H.
			\newblock Testing the Gravitational Field Generated by a Quantum Superposition.
			\newblock {\em New J. Phys.} {\bf 2019}, {\em 21},~093052.
			\newblock
			[\href{http://dx.doi.org/10.1088/1367-2630/ab41c1}{CrossRef}]
			
			\bibitem[Haine(2021)]{haine2021}
			Haine, S.A.
			\newblock Searching for Signatures of Quantum Gravity in Quantum Gases.
			\newblock {\em New J. Phys.} {\bf 2021}, {\em 23},~033020.
			\newblock
			[\href{http://dx.doi.org/10.1088/1367-2630/abd97d}{CrossRef}]
			
			\bibitem[Schm{\"o}le \em{et~al.}(2016)Schm{\"o}le, Dragosits, Hepach, and
			Aspelmeyer]{schmole2016}
			Schm{\"o}le, J.; Dragosits, M.; Hepach, H.; Aspelmeyer, M.
			\newblock A Micromechanical Proof-of-Principle Experiment for Measuring the
			Gravitational Force of Milligram Masses.
			\newblock {\em Class. Quantum Grav.} {\bf 2016}, {\em 33},~125031.
			\newblock
			[\href{http://dx.doi.org/10.1088/0264-9381/33/12/125031}{CrossRef}]
			
			\bibitem[Pedernales \em{et~al.}(2020)Pedernales, Morley, and
			Plenio]{pedernales2020a}
			Pedernales, J.S.; Morley, G.W.; Plenio, M.B.
			\newblock Motional {{Dynamical Decoupling}} for {{Matter-Wave Interferometry}}.
			\newblock {\em arXiv} {\bf 2020}, \href{https://doi.org/10.48550/arXiv.1906.00835}{arXiv:1906.00835}.
			
			\bibitem[Carney \em{et~al.}(2021)Carney, M{\"u}ller, and Taylor]{carney2021b}
			Carney, D.; M{\"u}ller, H.; Taylor, J.M.
			\newblock Using an {{Atom Interferometer}} to {{Infer Gravitational
					Entanglement Generation}}.
			\newblock {\em PRX Quantum} {\bf 2021}, {\em 2},~030330.
			\newblock
			[\href{http://dx.doi.org/10.1103/PRXQuantum.2.030330}{CrossRef}]
			
			\bibitem[Wiseman and Milburn(2009)]{wiseman2009}
			Wiseman, H.M.; Milburn, G.J.
			\newblock {\em Quantum Measurement and Control}; {Cambridge University Press}: Cambridge, UK,
			2009.
			
			\bibitem[Gross \em{et~al.}(2018)Gross, Caves, Milburn, and Combes]{gross2018}
			Gross, J.A.; Caves, C.M.; Milburn, G.J.; Combes, J.
			\newblock Qubit Models of Weak Continuous Measurements: Markovian Conditional
			and Open-System Dynamics.
			\newblock {\em Quantum Sci. Technol.} {\bf 2018}, {\em 3},~024005.
			\newblock
			[\href{http://dx.doi.org/10.1088/2058-9565/aaa39f}{CrossRef}]
			
			\bibitem[Gieseler \em{et~al.}(2020)Gieseler, Kabcenell, Rosenfeld, Schaefer,
			Safira, Schuetz, {Gonzalez-Ballestero}, Rusconi, {Romero-Isart}, and
			Lukin]{gieseler2020}
			Gieseler, J.; Kabcenell, A.; Rosenfeld, E.; Schaefer, J.D.; Safira, A.;
			Schuetz, M.J.A.; {Gonzalez-Ballestero}, C.; Rusconi, C.C.; {Romero-Isart},
			O.; Lukin, M.D.
			\newblock Single-{{Spin Magnetomechanics}} with {{Levitated Micromagnets}}.
			\newblock {\em Phys. Rev. Lett.} {\bf 2020}, {\em 124},~163604.
			\newblock
			[\href{http://dx.doi.org/10.1103/PhysRevLett.124.163604}{CrossRef}]
			
			\bibitem[Altamirano \em{et~al.}(2018)Altamirano, {Corona-Ugalde}, Mann, and
			Zych]{altamirano2018}
			Altamirano, N.; {Corona-Ugalde}, P.; Mann, R.B.; Zych, M.
			\newblock Gravity Is Not a Pairwise Local Classical Channel.
			\newblock {\em Class. Quantum Grav.} {\bf 2018}, {\em 35},~145005.
			\newblock
			[\href{http://dx.doi.org/10.1088/1361-6382/aac72f}{CrossRef}]
			
			\bibitem[Kovachy \em{et~al.}(2015)Kovachy, Asenbaum, Overstreet, Donnelly,
			Dickerson, Sugarbaker, Hogan, and Kasevich]{kovachy2015}
			Kovachy, T.; Asenbaum, P.; Overstreet, C.; Donnelly, C.A.; Dickerson, S.M.;
			Sugarbaker, A.; Hogan, J.M.; Kasevich, M.A.
			\newblock Quantum Superposition at the Half-Metre Scale.
			\newblock {\em Nature} {\bf 2015}, {\em 528},~530--533.
			\newblock
			[\href{http://dx.doi.org/10.1038/nature16155}{CrossRef}] [\href{http://www.ncbi.nlm.nih.gov/pubmed/26701053}{PubMed}]
			
			\bibitem[{Stamper-Kurn} \em{et~al.}(2016){Stamper-Kurn}, Marti, and
			M{\"u}ller]{stamper-kurn2016}
			{Stamper-Kurn}, D.M.; Marti, G.E.; M{\"u}ller, H.
			\newblock Verifying Quantum Superpositions at Metre Scales.
			\newblock {\em Nature} {\bf 2016}, {\em 537},~E1--E2.
			\newblock
			[\href{http://dx.doi.org/10.1038/nature19108}{CrossRef}] [\href{http://www.ncbi.nlm.nih.gov/pubmed/27582225}{PubMed}]
			
			\bibitem[Kovachy \em{et~al.}(2016)Kovachy, Asenbaum, Overstreet, Donnelly,
			Dickerson, Sugarbaker, Hogan, and Kasevich]{kovachy2016}
			Kovachy, T.; Asenbaum, P.; Overstreet, C.; Donnelly, C.A.; Dickerson, S.M.;
			Sugarbaker, A.; Hogan, J.M.; Kasevich, M.A.
			\newblock Kovachy et al. Reply.
			\newblock {\em Nature} {\bf 2016}, {\em 537},~E2--E3.
			\newblock
			[\href{http://dx.doi.org/10.1038/nature19109}{CrossRef}] [\href{http://www.ncbi.nlm.nih.gov/pubmed/27582226}{PubMed}]
			
			\bibitem[Cronenberg \em{et~al.}(2018)Cronenberg, Brax, Filter, Geltenbort,
			Jenke, Pignol, Pitschmann, Thalhammer, and Abele]{cronenberg2018}
			Cronenberg, G.; Brax, P.; Filter, H.; Geltenbort, P.; Jenke, T.; Pignol, G.;
			Pitschmann, M.; Thalhammer, M.; Abele, H.
			\newblock Acoustic {{Rabi}} Oscillations between Gravitational Quantum States
			and Impact on Symmetron Dark Energy.
			\newblock {\em Nat. Phys.} {\bf 2018}, {\em 14},~1022--1026.
			\newblock
			[\href{http://dx.doi.org/10.1038/s41567-018-0205-x}{CrossRef}]
			
			\bibitem[Schimmoller \em{et~al.}(2021)Schimmoller, McCaul, Abele, and
			Bondar]{schimmoller2021}
			Schimmoller, A.J.; McCaul, G.; Abele, H.; Bondar, D.I.
			\newblock Decoherence-Free Entropic Gravity: {{Model}} and Experimental Tests.
			\newblock {\em Phys. Rev. Res.} {\bf 2021}, {\em 3},~033065.
			\newblock
			[\href{http://dx.doi.org/10.1103/PhysRevResearch.3.033065}{CrossRef}]
			
			\bibitem[Verlinde(2011)]{verlinde2011}
			Verlinde, E.
			\newblock On the Origin of Gravity and the Laws of {{Newton}}.
			\newblock {\em J. High Energ. Phys.} {\bf 2011}, {\em 2011},~29.
			\newblock
			[\href{http://dx.doi.org/10.1007/JHEP04(2011)029}{CrossRef}]
			
			\bibitem[Hosten(2021)]{hosten2021}
			Hosten, O.
			\newblock Constraints on probing quantum coherence to infer gravitational entanglement.
			\newblock {\em Phys. Rev. Res.} {\bf 2021}, {\em 4},~013023.
			\newblock
			[\href{https://doi.org/10.1103/PhysRevResearch.4.013023}{CrossRef}]
			
			\bibitem[Ma \em{et~al.}(2021)Ma, Guff, Morley, Pikovski, and Kim]{ma2021}
			Ma, Y.; Guff, T.; Morley, G.; Pikovski, I.; Kim, M.S.
			\newblock Limits on Inference of Gravitational Entanglement.
			\newblock {\em Phys. Rev. Res.} {\bf 2021}, {\em 4},~013024.
			\newblock
			[\href{https://doi.org/10.1103/PhysRevResearch.4.013024}{CrossRef}]
			
			\bibitem[Carney \em{et~al.}(2021)Carney, Muller, and Taylor]{carney2021d}
			Carney, D.; Muller, H.; Taylor, J.M.
			\newblock Comment on ``{{Using}} an Atom Interferometer to Infer Gravitational
			Entanglement Generation''.
			\newblock {\em arXiv} {\bf 2021}, \href{https://doi.org/10.48550/arXiv.2111.04667}{arXiv:2111.04667}.
			
		\end{thebibliography}

\end{document}